\documentclass[prb,aps,twocolumn,showpacs,superscriptaddress,10pt]{revtex4-2}
\usepackage{amsfonts,amsmath,bm,multirow}
\usepackage[OT4]{fontenc}

\usepackage{graphicx}
\usepackage{xcolor}
\usepackage{epstopdf}
\usepackage{dsfont}
\usepackage{physics}
\usepackage{hyperref}
\usepackage{lipsum}
\usepackage{dcolumn}

\usepackage{natbib}
\usepackage{hf-tikz}

\definecolor{bred}{HTML}{e31a1c}
\definecolor{bgreen}{HTML}{33a02c}
\definecolor{bblue}{HTML}{1f78b4}

\definecolor{armygreen}{rgb}{0.29, 0.33, 0.13}

\newcolumntype{L}{D{.}{.}{3,4}}

\begin{document}

\title {Tunable \textit{g}-Factors of Hybridized Orbitals in a Quantum Dot Molecule}

\author{Michelle Lienhart$^{**}$}
\email{michelle.lienhart@tum.de}
\affiliation{Walter Schottky Institut, School of Natural Sciences, and MCQST, Technische Universität München, 85748 Garching, Germany}

\author{Krzysztof Gawarecki$^{**}$}
\email{krzysztof.gawarecki@pwr.edu.pl}
\affiliation{Institute of Theoretical Physics, Wroc\l{}aw University of Science and Technology, Wroc\l{}aw 50-370, Poland}

\author{Pavel Daskalov$^{**}$}
\affiliation{Walter Schottky Institut, School of Natural Sciences, and MCQST, Technische Universität München, 85748 Garching, Germany}

\author{Christopher Thalacker}
\affiliation{Walter Schottky Institut, School of Natural Sciences, and MCQST, Technische Universität München, 85748 Garching, Germany}

\author{Nadeem Akhlaq}
\affiliation{Walter Schottky Institut, School of Natural Sciences, and MCQST, Technische Universität München, 85748 Garching, Germany}

\author{Irina Ivanova}
\affiliation{Walter Schottky Institut, School of Natural Sciences, and MCQST, Technische Universität München, 85748 Garching, Germany}

\author{Johannes Schall}
\affiliation{Institut f\"ur Physik und Astronomie, Technische Universit\"at Berlin, Hardenbergstra\ss e 36, 10623 Berlin, Germany}

\author{Sven Rodt}
\affiliation{Institut f\"ur Physik und Astronomie, Technische Universit\"at Berlin, Hardenbergstra\ss e 36, 10623 Berlin, Germany}

\author{Stephan Reitzenstein}
\affiliation{Institut f\"ur Physik und Astronomie, Technische Universit\"at Berlin, Hardenbergstra\ss e 36, 10623 Berlin, Germany}

\author{Arne Ludwig}
\affiliation{Ruhr-University Bochum, Experimental Physics VI, Universit\"atsstra\ss e 150, 44801 Bochum, Germany}

\author{Dirk Reuter}
\affiliation{Paderborn University, Department of Physics, 33098 Paderborn, Germany}

\author{Kai M\"uller}
\affiliation{Walter Schottky Institut, School of Computation, Information and Technology, and MCQST, Technische Universität München, 85748 Garching, Germany}

\author{Jonathan J. Finley}
\email{jj.finley@tum.de \newline\newline$^{**}$ These authors contributed equally to this work.}
\affiliation{Walter Schottky Institut, School of Natural Sciences, and MCQST, Technische Universität München, 85748 Garching, Germany}

%%%%%%%%%%%%%%%%%%%%%%%%%%%%%%%%%%%%%%%
% Abstract
%%%%%%%%%%%%%%%%%%%%%%%%%%%%%%%%%%%%%%%

\begin{abstract}
The ability to control the $g$-factors of orbital spin states in optically active quantum dot molecules (QDMs) is a prerequisite for the high-fidelity generation of multi-photonic cluster states with higher-dimensional entanglement structure.
Protocols that rely on two coupled spins require knowledge of the $g$-factor and its dependence on external control parameters. Mismatches in the $g$-factor between tunnel-coupled dots introduce unwanted dephasing of coupled spin-states, making precise characterization and voltage control essential.
Here, we measure the gate voltage dependence of the electron and hole $g$-factors of negatively charged trions $X^{-}$ in a single InGaAs QDM using polarization-resolved magneto-photoluminescence spectroscopy. The electron $g$-factor exhibits a pronounced step-like change at the tunneling resonance, shifting from $g_\mathrm{e} = -0.336\pm 0.008$ to $g_\mathrm{e} = -0.389\pm 0.003$, providing a direct spectroscopic fingerprint of molecular orbital formation and a shift of the wavefunction localization from the lower to the upper dot. In contrast, the hole $g$-factor remains nearly constant at $g_\mathrm{h} \approx 0.094 \pm 0.007$, exhibiting a weak modulation near the anticrossing voltages attributed to Coulomb-mediated deformation of the wavefunction by the tunneling electron.
Our results are quantitatively reproduced by an eight-band $\mathbf{k}{\cdot}\mathbf{p}$ model, establishing electric-field control of the trion $g$-factors as a practical tool for independently tuning the Zeeman splitting of individual dots and opening new pathways towards the deterministic generation of two-dimensional photonic cluster states.
\end{abstract}

\maketitle

%%%%%%%%%%%%%%%%%%%%%%%%%%%%%%%%%%%%%%%
% Introdcution
%%%%%%%%%%%%%%%%%%%%%%%%%%%%%%%%%%%%%%%

\section{Introduction}
\label{sec:intr}

Optically active semiconductor quantum dots (QDs) have emerged as near ideal sources for quantum photonic technologies~\cite{Heindel2023,Huber2026,Lodahl2018}. The strong optical activity and well-defined polarization selection rules of their spin states make them suitable for entangling sequentially emitted photons to form for instance photonic cluster states~\cite{Loss1998,Vrijen2000,Press2008,Kim2008}.  Beyond that, pairs of vertically stacked QDs, quantum dot molecules (QDMs), have orbital states that can be hybridized via a judicious choice of the static electric field, controlled via a gate voltage \cite{Krenner2005}. They can be populated by interacting spins that can be optically addressed to produce entangled few-photon states with multi-dimensional entanglement structures needed for measurement-based quantum information processing and communication \cite{economouOpticallyGenerated2Dimensional2010,vezvaee2022,lindnerProposalPulsedOnDemand2009,Cogan2023,Schwartz2025, Gimeno2019}. The routes toward such applications critically depend on the ability to control and maintain spin-coherence, limited by hyperfine interaction with the surrounding nuclear spin bath~\cite{Merkulov2002,Khaetskii2002,Hanson2007}, as well as spin-orbit ~\cite{Golovach2004,Bulaev2005} and phonon-mediated spin and orbital relaxation~\cite{Lienhart2025, nakaokaDirectObservationAcoustic2006}.

Enabling complete coherent spin control requires the use of an external magnetic field $B$. This field splits the energy states according to the Zeeman energy $\Delta E = g \mu_B B$, where $\mu_B$ is the Bohr magneton and $g$ is the Landé $g$-factor, which is the fundamental parameter that governs the response of spins to external magnetic fields.
The $g$-factor value in QDs differs substantially from its bulk counterpart~\cite{Pryor2006}, as the quantum confinement affects the spin-correlated orbital currents~\cite{VanBree2016}. The ability to tune the $g$-factor and its tensor elements is of great importance for spin-based quantum technologies. For example, it has been suggested to facilitate all electrical control over the spin wavefunction \cite{Pryor2006, Doty2006}.  Moreover, an electron $g$-factor close to zero suppresses Zeeman splitting between spin states in electron-storage samples, and has been predicted to enhance the performance of quantum repeaters~\cite{Kosaka2003, Zajac2025}. Conversely, a large hole $g$-factor breaks the valence-band degeneracy, facilitating efficient spin initialization and reliable entanglement fidelity~\cite{Vrijen2001}. Previous work has demonstrated that mismatches in $g$-factors between tunnel-coupled QDs give rise to differential evolution of the phase of the spin wavefunction, resulting in pure dephasing~\cite{Gawelczyk2018} that poses challenges for the deterministic generation of photonic cluster states~\cite{economouOpticallyGenerated2Dimensional2010}.

The $g$-factors in single self-assembled QDs have been extensively investigated, both experimentally and theoretically. On the experimental side, measurements of electron and hole $g$-factors have been performed in various material systems~\cite{Nakaoka2004,Gawarecki2018a,Bayer1999,Jovanov2012,Jovanov2011,Kleemans2009}, demonstrating strong dependencies on dot size, shape, composition, and strain. Theoretically, these observations are well captured by multi-band $\mathbf{k}{\cdot}\mathbf{p}$ models~\cite{MedeirosRibeiro2003} as well as atomistic tight-binding approaches~\cite{Gawarecki2025,Nakaoka2004,Jovanov2012,Nakaoka2005}.

An important early advance was made by Doty \textit{et al.}, who showed that $g$-factors in QDMs can be electrically tuned by controlling the hybridization of coupled orbitals~\cite{Doty2006}. By applying an electric field to tune the relative energies of the orbital states in the two QDs, they observed strong resonant changes in the hole $g$-factor for the neutral exciton $X^0$ and the positively charged trion $X^{+}$. These effects were shown to arise from the formation of bonding and antibonding hole orbitals that modulate the amplitude of the carrier wavefunction in the tunnel barrier. Crucially, the electron $g$-factor showed no measurable electric-field dependence in their samples, since the electron remained localized in the bottom dot throughout the studied field range.
Liu \textit{et al.} introduced an Al$_{0.3}$Ga$_{0.7}$As layer into the tunnel barrier of an InAs QDM, exploiting the strongly composition-dependent electron $g$-factor and opposite sign of the AlGaAs $g$-factor to enhance tunability~\cite{Liu2011}. By tuning the electron levels into resonance via an applied electric field, they demonstrated a $50\%$ in situ change in the excitonic $g$-factor, confirmed through measurements of both the neutral exciton $X^0$ and the doubly negatively charged exciton $X^{2-}$. However, the electron and hole $g$-factors could not be independently resolved in their experiment, leaving the individual spin contributions uncharacterized. Complementary theoretical work by Andlauer and Vogl quantitatively reproduced the resonant $g$-factor enhancements using an eight-band $\mathbf{k}{\cdot}\mathbf{p}$ framework that incorporates the magnetic field in a gauge-invariant manner, and further predicted a giant electrically tunable anisotropy of hole $g$-factors due to piezoelectric charges~\cite{Andlauer2009}. Together, these studies established QDMs as a uniquely versatile system in which the sign and magnitude of spin $g$-factors can be manipulated using static electric fields.

Despite this progress, a systematic experimental characterization of the trion ($X^{-}$), electron, and hole $g$-factors in a QDM across the full voltage tuning range, spanning the tunneling resonance where the electron wavefunction shifts between dots, has not been reported. Knowledge of the trion $g$-factors has recently become relevant in the context of spin initialization and coherent spin control protocols required for cluster-state generation~\cite{economouOpticallyGenerated2Dimensional2010,Cogan2023, Gimeno2019}: the charged exciton couples the spin degree of freedom to the optical field, and its Zeeman structure sets the conditions for optical spin pumping, spin-selective excitation, and entangled photon generation. Furthermore, as already alluded to, mismatches between the electron g-factors of the two constituent dots introduce a pure dephasing channel that limits the attainable fidelities of two-qubit spin gates~\cite{Gawelczyk2018}, making precise voltage control of this parameter indispensable for applications.

In this work, we measure both the electron and hole $g$-factors of the $X^{-}$ state in a single InGaAs QDM over a large electric field range. Our data explicitly resolve the transfer of the electron between the two dots as a pronounced and characteristic change in the measured electron $g$-factor, providing a direct spectroscopic fingerprint of molecular orbital formation. We accompany these measurements with eight-band $\mathbf{k}{\cdot}\mathbf{p}$ theory, which accurately reproduces the observed voltage dependence.
The reliability of the theoretical model is further validated by photoluminescence voltage-sweep (PLV) measurements, which demonstrate that the simulated QDM energy levels are in very good quantitative agreement with the experiment.
Unlike previous studies that focused on neutral exciton $g$-factors near the tunneling resonance~\cite{Doty2006,Andlauer2009}, our work with a charge-storage device targets the charged trion throughout the full bias range, provides separate resolution of electron and hole $g$-factors, and demonstrates the quantitative predictive power of the theoretical framework across all accessible few-particle configurations.
Our results firmly establish electric-field control of the trion $g$-factors as a practical and precisely characterizable tool for independently tuning the Zeeman splitting of individual QDs, opening new pathways towards the deterministic generation of high-fidelity photonic cluster states from coupled quantum dot systems.

The paper is structured as follows: In Sec.\,\ref{sec:experiment}, we describe the sample design, introduce the experimental methods, and present results for the electron and hole $g$-factors over the full voltage tuning range. In Sec.\,\ref{sec:theory}, we define the theoretical model used to quantitatively interpret the experimental results and understand the microscopic cause of the observed electric field tunabilities. Finally, Sec.\,\ref{sec:conclusions} compares experiment with theory and summarizes the major conclusions of our work. Additional details on the sample structure, experimental data analysis, and theoretical calculations are provided in the Appendix.

%%%%%%%%%%%%%%%%%%%%%%%%%%%%%%%%%%%%%%%
% Experiment
%%%%%%%%%%%%%%%%%%%%%%%%%%%%%%%%%%%%%%%

\section{Experiment}
\label{sec:experiment}

In this section, we begin by describing the sample structure and experimental methods before moving on to present the electric field dependent magneto-optical data.

\subsection{Photoluminescence spectrum and level structure}
\begin{figure}[tb]
    \centering
    \hspace{0.02cm}\includegraphics[width=0.995\linewidth]{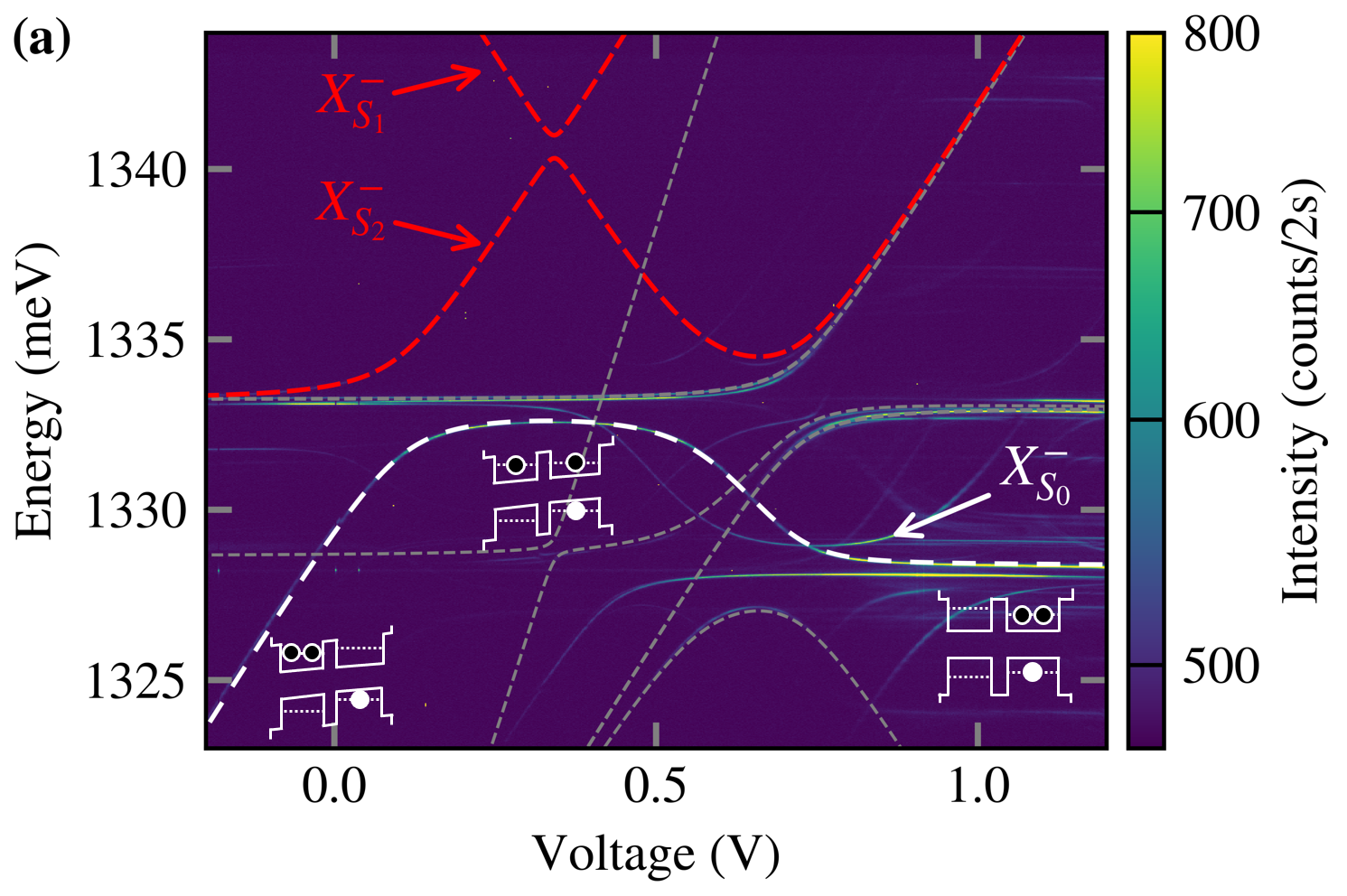}
     \hspace{-0.25em}\raisebox{1em}{\includegraphics[width=0.431\linewidth]{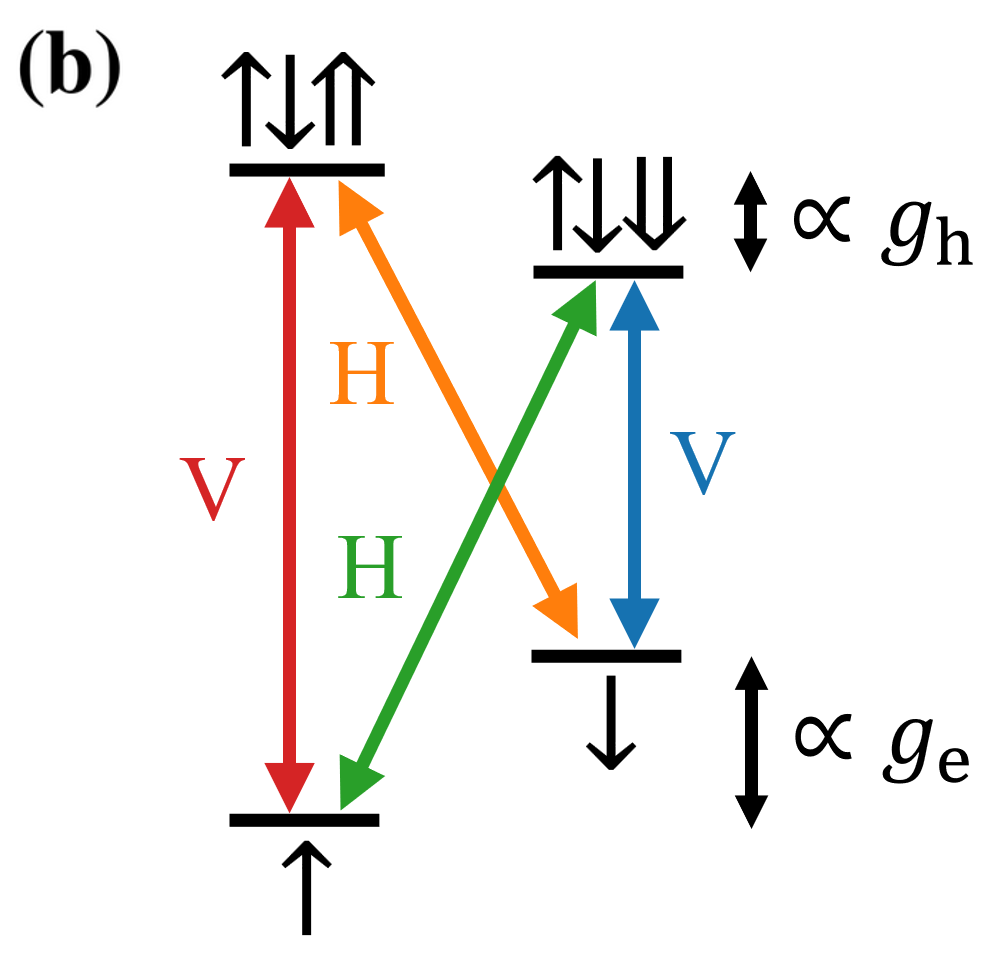}}
    \hspace{0.1em}
    \includegraphics[width=0.518\linewidth]{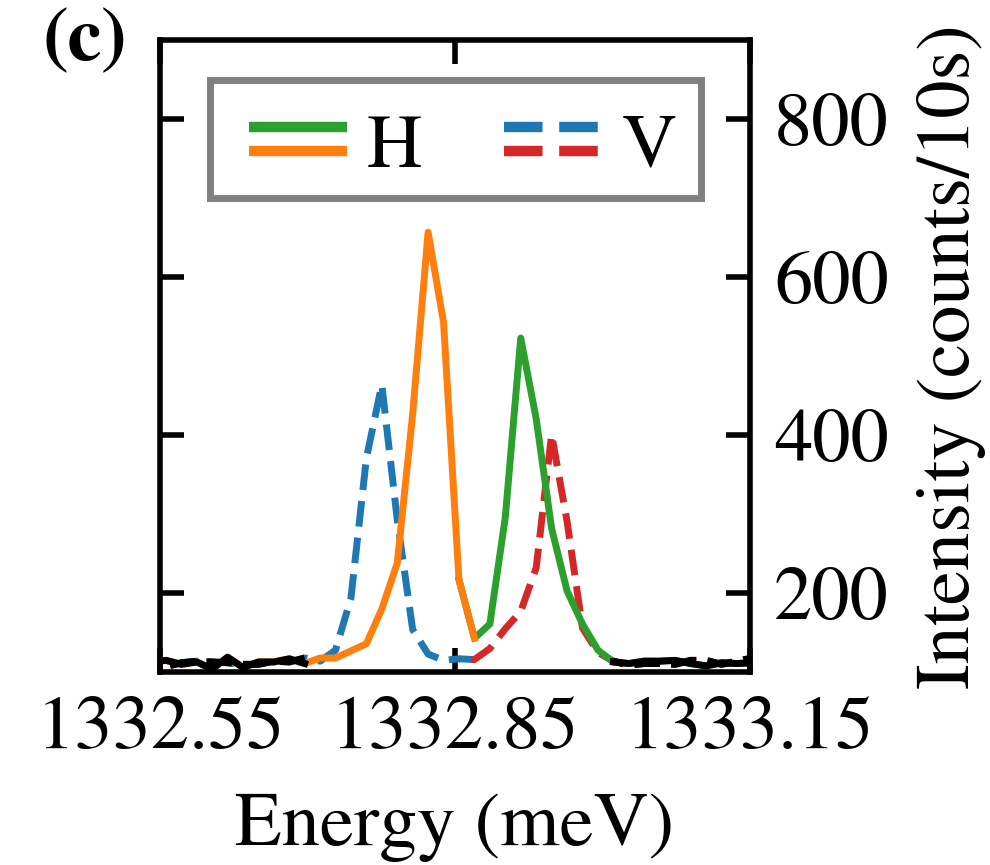}
    \caption{(a) Voltage-dependent photoluminescence recorded from the single QDM investigated by exciting quasi-resonantly in the wetting layer. The trion states are fitted using a few-body Hamiltonian. The relevant singlet states are highlighted by the white curve ($X^{-}_{S_0}$) and red curves ($X^{-}_{S_1}$ and $X^{-}_{S_2}$). The insets schematically depict the voltage dependent occupation of the orbital states in the trion singlet $X^{-}_{S_0}$. (b) Schematic representation of the trion level structure when applying an in-plane magnetic field (Voigt geometry). (c) Polarization-resolved emission from the trion  $X^{-}_{S_0}$ state at $B=7\,$T and 0.44\,V.}
    \label{fig:PLV_trionlevels}
\end{figure}

The quantum dot molecule (QDM) investigated in this work consists of two vertically stacked InGaAs quantum dots separated by $\approx10$\,nm \cite{Lienhart2025}. The quantum dots were grown using the In-flush technique \cite{Fafard1999}, which enables precise control over their respective heights. The QDM is embedded into the intrinsic region of a p-i-n diode, allowing the application of a gate voltage to tune the electric field along the growth axis.
The layer structure is engineered to store electrons within the QDM and at the same time tune the electron orbitals into resonance: the lower QD is intentionally flatter, providing a larger confinement energy than the upper dot. A 50\,nm thick $\mathrm{Al_{0.33}Ga_{0.67}As}$ barrier below the QDMs and the n-contact suppresses electron tunneling out of the QDM. Due to the geometry, the hole is always located in the upper dot in our experiments. An additional 2.5\,nm thick $\mathrm{Al_{0.33}Ga_{0.67}As}$ layer between the dots acts as a tunnel barrier, reducing the strength of the interdot coupling.
A circular Bragg grating is deterministically fabricated in the p-doped layer directly above the QDM using in-situ electron beam lithography \cite{Schall2021}. Together with a distributed Bragg reflector beneath the QDM, this structure enhances photon-collection efficiency over a bandwidth of approximately 10\,nm, broad enough to address optical transitions in both the upper and lower quantum dots \cite{Schall2021}.
Further information on the epitaxial growth and device fabrication can be found in Appendix\,\ref{app:sample} and Refs.\,\cite{Bopp2023a, BoppMagTun, Lienhart2025, Thalacker2026}.

Figure\,\ref{fig:PLV_trionlevels}(a) presents typical voltage-dependent photoluminescence spectroscopy (PLV) recorded from the investigated QDM under non-resonant wetting layer (850\,nm) excitation. Under these conditions, fluctuations in the charge occupation of the QDM enable simultaneous observation of emission from multiple charge states.
We make use of a periodic, two-step electrical and optical measurement protocol: First, we remove all charges from the QDM by applying a large negative gate voltage of -10\,V (reset phase) for a time $t_{\mathrm{reset}}$. In a following step, the gate voltage is tuned to manipulate the relative energy separation of the electronic orbital states in each of the two dots as the system is optically excited for a time $t_{\mathrm{read}}$ (readout phase) \cite{boppQuantumDotMolecule2022}.  The data presented in Fig.\,\ref{fig:PLV_trionlevels}(a) were recorded with $T=t_{\mathrm{reset}}+t_{\mathrm{read}}=2.56\,\mu$s and a duty cycle $t_{\mathrm{read}}/T$ of 65\%.

The different charge states visible in Fig.\,\ref{fig:PLV_trionlevels}(a) are identified by fitting the data with a few-body Hamiltonian~\cite{Thalacker2026}. For clarity, only the charge states belonging to the negatively charged trion $X^{-}$ are highlighted. 
This section will focus on the singlet line $X^{-}_{S_0}$ shown in white. The singlet lines $X^{-}_{S_1}$ and $X^{-}_{S_2}$ (shown in red) will be discussed in Sec.\,\ref{sec:theorymodel}.

The $g$-factor is determined from measurements of the trion singlet line $X^{-}_{S_0}$. 
Due to the sample geometry, the hole remains localized in the upper dot throughout the entire voltage range investigated.  The charge configuration insets in Fig.\,\ref{fig:PLV_trionlevels}(a) depict the electron and hole distribution among the two dots forming the molecule in different regimes. At large gate voltages ($> 0.7$\,V), the two electrons in the initial trion state are primarily confined in the upper QD, analogous to the situation in a single dot. The trion decays to the final state in which a single electron remains in the upper dot.
As the gate voltage in the readout phase is progressively reduced below $\sim0.7$\,V, electrons in the excited trion state hybridize across both dots, resulting in a configuration with one electron in the upper and one electron in the lower dot near $0.4$\,V. Further reduction of the gate voltage below $\sim0.2$\,V results in both electrons becoming fully localized in the lower dot.

In an in-plane magnetic field (Voigt geometry) the split trion states and the split single electron states form a double-$\Lambda$ system, as shown in Fig.\,\ref{fig:PLV_trionlevels}(b)~\cite{Bayer2002,Gywat2009,Knight1997}. A key feature of the Voigt geometry is that the in-plane magnetic field results in all four trion transitions becoming optically allowed.
These transitions naturally group into two sets of orthogonal linear polarizations, labeled horizontal (H) and vertical (V), with energy splittings proportional to the electron and hole $g$-factors ($g_\mathrm{e}$ and $g_\mathrm{h}$, respectively). The resulting well-defined optical selection rules form the basis for spin-photon interface experiments, since they enable controlled addressing and readout of individual spin states via polarized optical pulses \cite{Press2008}.
Fig.\,\ref{fig:PLV_trionlevels}(c) shows the four linearly polarized trion transitions measured at $B = 7\,$T and a gate voltage of $0.44\,$V in the hybridized electron regime.

For measuring the $g$-factor of the trion singlet emission we apply a magnetic field and then utilize an electrical and optical sequence identical to the one used to produce Fig.\,\ref{fig:PLV_trionlevels}(a). At each magnetic field two PLV measurements are done, once detecting for H-polarized emission and once detecting for V-polarized emission, giving us access to the energy splittings at the different voltages and different magnetic fields. Two example polarization-resolved PLV measurements taken at $B=8$\,T can be found in Appendix\,\ref{app:setup} in Fig.\,\ref{fig:PLV_H_V}.

\subsection{Electron and hole \textit{g}-factors}
\label{sec:gfactor_experiment}

\begin{figure}[tb]
    \centering
    \hspace{0.5em}
    \includegraphics[width=0.99\linewidth]{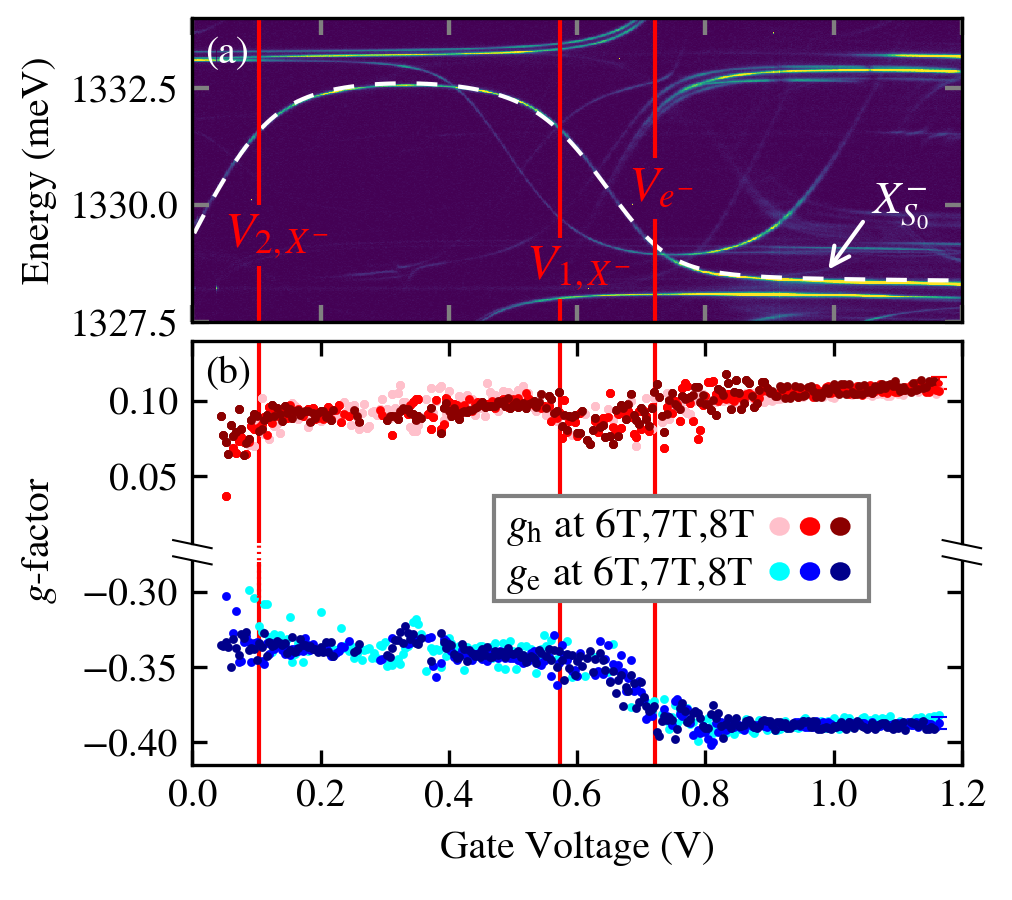}
    \includegraphics[width=1.01\linewidth]{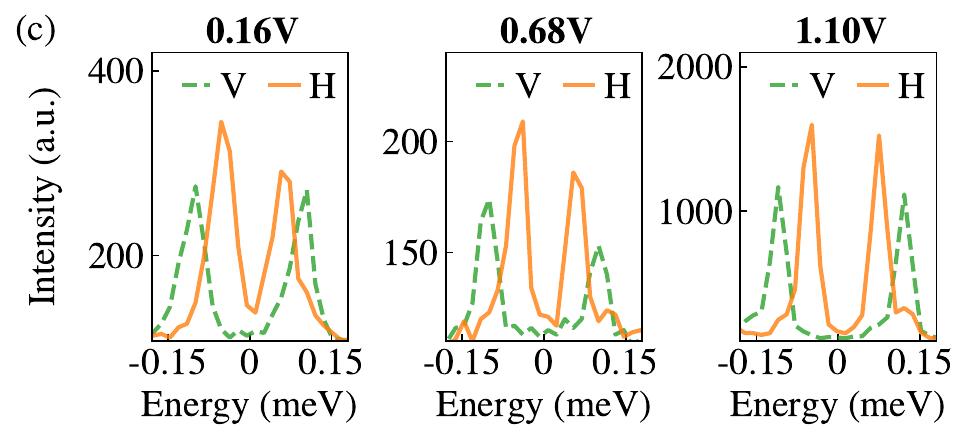}
    \caption{(a) Voltage-dependent photoluminescence recorded from the trion singlet state $X^{-}_{S_0}$ with the trion $V_{1,X^{-}}$, $V_{2,X^{-}}$ and electron $V_{e^{-}}$ avoided crossings shown by vertical red lines.
    (b) Voltage dependent electron $g_{\mathrm{e}}$ (blue) and hole $g_{\mathrm{h}}$ (red) $g$-factors showing no B-field dependence. The electron $g$-factor $g_{\mathrm{e}}$ changes around the electron avoided crossing $V_{e^{-}}$.
    (c) Polarization-resolved emission from the trion $X^{-}_{S_0}$ state at 8\,T for the voltages 0.16\,V, 0.68\,V, and 1.1\,V. The zero of the energy axis is set approximately at the midpoint of the splitting. The corresponding PLV can be found in the Appendix\,\ref{app:setup} in Fig.\,\ref{fig:PLV_H_V}.
    }
    \label{fig:gfactor_Xm_Voigt}
\end{figure}

The electron and hole $g$-factors are extracted from PLV measurements recorded at three magnetic fields ($6\,$T, $7\,$T, and $8\,$T). At each voltage, the peak positions of the $X^{-}_{S_0}$ transitions are obtained by fitting Lorentzian peaks to the trion emission in each spectral slice of these PLV measurements (representative spectra are shown in Fig.\,\ref{fig:PLV_trionlevels}(c) and Fig.\,\ref{fig:gfactor_Xm_Voigt}(c)). The energy splitting of each transition pair $\Delta E_{\mathrm{H}/\mathrm{V}}$ yields the $g$-factors for the H- and V-polarized spectra via
\begin{equation}
    g_{H/V} = \frac{|\Delta E_{\mathrm{H}/\mathrm{V}}|}{\mu_\mathrm{B} B},
\end{equation}
where $B$ is the applied magnetic field and $\mu_\mathrm{B}$ is the Bohr magneton. The electron and hole $g$-factors are then given by $g_\mathrm{e} = |g_{\mathrm{H}} + g_{\mathrm{V}}|/2$ and $g_\mathrm{h} = |g_{\mathrm{H}} - g_{\mathrm{V}}|/2$~\cite{Bayer2002}, respectively.

Fig.\,\ref{fig:gfactor_Xm_Voigt}(a) presents the voltage-dependent photoluminescence of the $X^{-}_{S_0}$ state (white line), with the trion anticrossing voltages $V_{1,X^{-}}$ and $V_{2,X^{-}}$ and the electron anticrossing voltage $V_{e^{-}}$ indicated by vertical red lines. $V_{e^{-}}$ denotes the voltage at which the final-state electron experiences a tunneling resonance from the upper dot to the lower dot. $V_{1,X^{-}}$ and $V_{2,X^{-}}$ mark the voltages at which each of the two trion-state electrons successively tunnel from the upper dot to the lower dot.

Fig.\,\ref{fig:gfactor_Xm_Voigt}(c) shows examples of the polarization-resolved emission from the $X^{-}_{S_0}$ state at 8\,T for three voltages 0.16\,V, 0.68\,V, and 1.1\,V from which the $g$-factors were extracted.
The voltage-dependent $g_\mathrm{e}$ and $g_\mathrm{h}$ values resulting from all spectra are summarized in the broken-axis graph in Fig.\,\ref{fig:gfactor_Xm_Voigt}(b) for all three magnetic fields.
$g_\mathrm{e}$ exhibits a clear voltage-dependent step around $V_{e^{-}}$, reducing from $-0.336\pm0.008$ (low voltage) to $-0.389\pm0.003$ (high voltage), which directly reflects the transfer of the final-state electron from the lower to the upper dot. 
In contrast, $g_\mathrm{h}$ remains comparatively flat, showing only a slight increase from $0.078\pm0.007$ to $0.110\pm0.004$ from low to high voltages with a shallow dip near the anticrossings $V_{1,X^{-}}$ and $V_{e^{-}}$. 
For visual clarity, individual error bars have been omitted, except for two high-voltage data points from the 7T dataset. The overall measurement uncertainty, attributed to the finite integration time and spectrometer resolution limiting the fit accuracy, can otherwise be inferred from the spread of the data. 
Notably, neither $g_\mathrm{e}$ nor $g_\mathrm{h}$ shows any measurable magnetic-field dependence across the three fields studied.

The observed behavior can be understood in terms of the electron and hole wavefunctions sampling different local environments as the electric field (voltage) is varied. The strain profile, quantum confinement strength, and material composition differ between the two dots, resulting in dot-specific $g$-factor values~\cite{Jovanov2011,Nakaoka2004}. As the electron wavefunction shifts from the lower to the upper dot it experiences their specific environment characteristics, giving rise to the step-like feature observed in $g_\mathrm{e}$ around $V_{e^{-}}$. This is discussed in further detail under Sec.\,\ref{sec:theoryLande}.

Although the hole does not tunnel between QDs, its $g$-factor $g_\mathrm{h}$ nevertheless exhibits a shallow dip near the $V_{e^{-}}$ and $V_{1,X^{-}}$ anticrossings. This is attributed to a deformation of the hole wavefunction inside the upper dot driven by Coulomb coupling to the tunneling electron in the initial ($X^-$) state. A related mechanism was previously identified by Doty \textit{et al.}~\cite{Doty2006}, where resonant $g$-factor changes arose from the hole wavefunction sampling the tunnel barrier during hole tunneling. In the present QDM the hole remains confined to the upper dot, yet still exhibits a weak $g$-factor modulation through this indirect coupling. The weak shift of $g_\mathrm{h}$ to higher values with increase in voltage is attributed to the quantum-confined Stark shift, consistent with previous observations~\cite{Jovanov2011}.

%%%%%%%%%%%%%%%%%%%%%%%%%%%%%%%%%%%%%%%
% Theory
%%%%%%%%%%%%%%%%%%%%%%%%%%%%%%%%%%%%%%%

\section{Theory}
\label{sec:theory}

We continue by defining the theoretical approaches and models used to calculate the voltage dependent spectrum of the excitonic complexes and present the results of our simulations.

\subsection{Model}
\label{sec:theorymodel}
\begin{figure}[tb]
    \centering
    \includegraphics[width=0.9\linewidth]{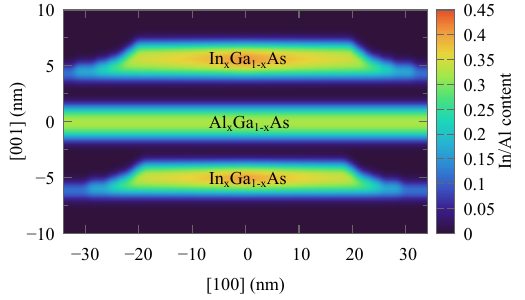}
    \caption{The In content distribution (In$_{x}$Ga$_{1-x}$As) in the QDs and Al content (Al$_{x}$Ga$_{1-x}$As) in the interdot barrier.}
    \label{fig:qd_geometry_theory}
\end{figure}

As described in Ref.~\cite{Lienhart2025}, the QDs are modeled using a truncated-Gaussian shape (see Fig.~\ref{fig:qd_geometry_theory}), with the trumpet-shaped composition profile~\cite{Jovanov2012,Migliorato2002}, well established for InGaAs QDs. The model parameters are provided in Appendix~\ref{app:theory}.

The lattice mismatches between InGaAs, GaAs, and AlGaAs generate strain that strongly affects the system properties. We account for this by calculating the strain distribution using the continuum elasticity approach~\cite{Pryor1998b}, and include piezoelectric fields up to second-order strain-induced polarization~\cite{Bester2006b,Caro2015}.

The electron and hole states are calculated within the eight-band $\mathbf{k}\cdot\mathbf{p}$ model using the envelope function approximation~\cite{Trebin1979,Winkler2003}; implementation details are given in Refs.~\cite{Gawarecki2018a,Lienhart2025}. The magnetic field is incorporated via a gauge-invariant scheme~\cite{Andlauer2008}.

The single-particle wave functions form the basis for calculating the negative trion $X^-$ states using the configuration interaction (CI) method. Radiative transitions between the $X^-$ and single-electron final states are calculated within the dipole approximation~\cite{Haug2004,Thranhardt2002}, as described in Refs.~\cite{Gawarecki2023,Gawarecki2025}, and spectral lines are broadened using Gaussian functions.

The trion energy spectrum grows rapidly in complexity with the CI basis size. In practice, however, most excited $X^-$ configurations do not contribute to the emission spectrum, as they undergo fast non-radiative relaxation to lower-lying states. We therefore restrict the calculations to the $s$-shell single-particle manifold, in which the CI space consists of four electron states (two $s$-like states for each spin projection in each QD) and two hole states (the $s$-like states in the upper QD).

\subsection{Results: energy levels and radiative transitions}

\begin{figure}[tb]
    \centering
    \includegraphics[width=\linewidth]{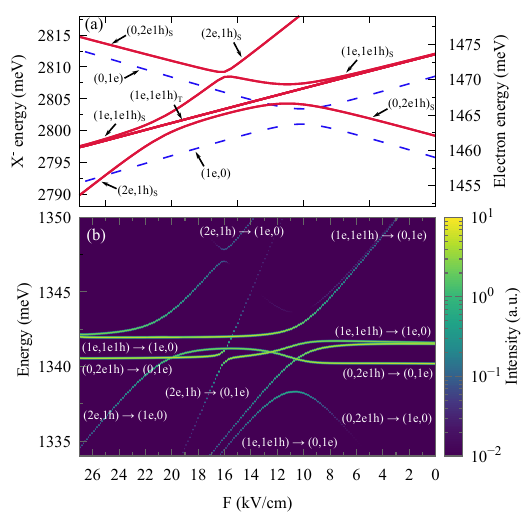}
    \caption{(a) Calculated negative trion $X^-$ energies (red solid lines, scale on the left) and single-electron energy branches (blue dashed lines, scale on the right), and (b) the simulated PLV spectrum.}
    \label{fig:plv_theory}
\end{figure}

We begin by focusing on the single-electron energy levels. As shown in Fig.~\ref{fig:plv_theory}(a) the energy branches for the electron in the lower dot (1e,0) and in the upper dot (0,1e) have an avoided crossing characterized by an energy $\Delta E_\mathrm{ac} \approx 2.2$~meV at $F \approx 10.2$~kV/cm. It is a manifestation of the well-known tunnel resonance, where the electron changes its localization between the two dots~\cite{Krenner2005,Krenner2006}.  This resonance is the theoretical counterpart of the experimentally defined electron anticrossing voltage $V_{e^-}$ in Fig.~\ref{fig:gfactor_Xm_Voigt}(a).

We also calculated the energy branches for the negative trion $X^-$ (red solid lines in Fig.~\ref{fig:plv_theory}(a)). As two electrons are now present, they form spin-singlet and spin-triplet configurations. The triplet state (1e,1e1h)$_\mathrm{T}$ is decoupled from the singlet states, and its energy shows a Stark shift with a slope very similar to that of the single-electron case (1e,0). On the other hand, the singlet states (2e,1h)$_\mathrm{S}$ and (0,2e1h)$_\mathrm{S}$ exhibit anticrossings with (1e,1e1h)$_\mathrm{S}$ at $F \approx 20.7$~kV/cm and  $F \approx 11.2$~kV/cm, respectively. These two singlet resonances correspond to the experimentally labeled
trion anticrossing voltages $V_{2,X^-}$ and $V_{1,X^-}$, respectively. Their widths amount to $\Delta E_\mathrm{ac} \approx 3.0$~meV which, as expected~\cite{Schreibner2007}, is approximately $\sqrt{2}$ times larger than the bare single-electron value. 
Finally, at $F \approx 16$~kV/cm, there is an avoided crossing of (2e,1h)$_\mathrm{S}$ and (0,2e1h)$_\mathrm{S}$ branches. Its width is much smaller ($\Delta E_\mathrm{ac} \approx 0.8$~meV) owing to the two-particle character of this resonance.

Fig.~\ref{fig:plv_theory}(b) shows the results of our calculations of the emission spectrum associated with recombination of the negative trion into the single‑electron final states. This can be compared with the experimental PLV map shown in Fig.~\ref{fig:PLV_trionlevels}(a). We first focus on the transitions, which are the theoretical counterpart of the experimentally followed $X^-_{S_0}$ line. 
The lowest-energy line at $F = 27$~kV/cm corresponds to the transition from the singlet (2e,1h)$_\mathrm{S}$ to the (1e,0) final state, i.e. recombination of an electron and hole localized in different dots. 
Due to the above-mentioned (2e,1h)$_\mathrm{S}$--(1e,1e1h)$_\mathrm{S}$ avoided crossing in the initial state,  for electric fields from $20.7$~kV/cm to $11.2$~kV/cm the line is associated with the transition (1e,1e1h)$_\mathrm{S} \rightarrow$ (1e,0). 
In that case, the recombining electron-hole pair is confined within the same QD, therefore the emission exhibits a high intensity. Finally, after the (1e,1e1h)$_\mathrm{S}$--(0,2e1h)$_\mathrm{S}$ avoided crossing at $F \approx 11.2$~kV/cm and the (1e,0)--(0,1e) one at $F \approx 10.2$~kV/cm, the line describes a pronounced transition (0,2e1h)$_\mathrm{S} \rightarrow$ (0,1e). 

In addition to the fundamental transition, the spectrum contains a variety of additional features. The horizontal line which starts from the energy of $1342$~meV at $F = 27$~kV/cm corresponds to the (1e,1e1h)$_\mathrm{T} \rightarrow$ (1e,0) recombination. While the initial triplet state remains in the (1e,1e1h)$_\mathrm{T}$ configuration, the final state exhibits an anticrossing (cf. $V_{e^-}$), which leads to a change in the line slope from $F \approx 10.2$~kV/cm. 

The spectrum also contains lines that are associated with the radiative Auger transitions. For example the (0,2e1h)$_\mathrm{S} \rightarrow$ (1e,0) describes the process, where the electron-hole recombination in the upper QD is accompanied by the promotion of the remaining electron to the state (1e,0) in the lower dot. Therefore, the emission energy is redshifted with respect to the (0,2e1h)$_\mathrm{S} \rightarrow$ (0,1e) case. While the strength of radiative Auger processes is much weaker than that of the usual radiative transitions, their presence is revealed by avoided crossings visible in the spectrum. For instance, at $F \approx 16$~kV/cm, the upper (in the energy) anticrossing takes place between the branches representing the (2e,1h)$_\mathrm{S} \rightarrow$ (1e,0) and the (0,2e1h)$_\mathrm{S} \rightarrow$ (1e,0) transitions; the energetically lower anticrossing appears between (0,2e1h)$_\mathrm{S} \rightarrow$ (0,1e) and the (2e,1h)$_\mathrm{S} \rightarrow$ (0,1e) lines. Both of them have spectral fingerprints arising from the two-electron tunnel resonance discussed above. The Auger-related branches hybridize with the ordinary radiative transitions and contribute to the
pattern observed in the experimental trion spectrum in Fig.~\ref{fig:PLV_trionlevels}(a), e.g. in lines $X^-_{S_1}$ and $X^-_{S_2}$.

The theoretical results reproduce the experimental data from Fig.~\ref{fig:PLV_trionlevels}(a) very well: the spectral features and avoided crossing patterns are faithfully captured, and both experiment and theory consistently predict the existence of Auger transitions. The energy splittings are also in good quantitative agreement. Specifically, the measured splitting at the avoided crossing between the singlet configurations (2e,1h)$_\mathrm{S}$--(1e,1e1h)$_\mathrm{S}$ (at $0.051$~V) is $\Delta_\mathrm{ac} = 2.95$~meV, compared to the theoretical value of $3.0$~meV. The singlet-triplet splitting (at $0.17$~V) is $0.655$~meV experimentally, while theory predicts $0.79$~meV. Finally, the tunnel coupling strength at the two-electron avoided crossing (2e,1h)$_\mathrm{S}$--(0,2e1h)$_\mathrm{S}$ is $0.676$~meV in experiment and $0.80$~meV in theory.

\subsection{Land\'{e} \textit{g}-factor results}
\label{sec:theoryLande}
\begin{figure}[tb]
    \centering
    \includegraphics[width=\linewidth]{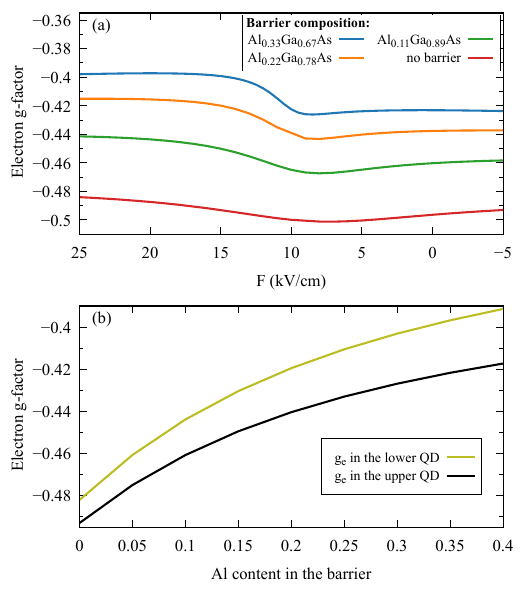}
    \caption{(a) The electron $g$-factor as a function of the electric field, calculated  for various compositions of the AlGaAs barrier.  (b) The $g$-factor at $F = -5$~kV/cm and $F = 25$~kV/cm associated with the values in the upper and the lower QD, respectively.}
    \label{fig:ge_theory}
\end{figure}

The electron $g$-factor as a function of electric field is presented in Fig.~\ref{fig:ge_theory}(a). We first consider the results obtained for the Al$_{0.33}$Ga$_{0.67}$As barrier between the QDs, matching the experimentally studied sample. Consistent with the measurements, the calculations reveal a pronounced step associated with electron tunneling between the QDs. As seen in Fig.~\ref{fig:plv_theory}(a), this step aligns with the corresponding single-electron energy avoided crossing, confirming the origin of the feature observed in our experiments. The calculated electron $g$-factors for the upper and lower QDs are $-0.424$ and $-0.398$, respectively, in good agreement with the measured values of $-0.389\pm 0.003$ and $-0.336\pm 0.008$.

To gain insight into the mechanisms responsible for the $g$-factor difference between the two QDs, we compare these results to the case where the Al$_x$Ga$_{1-x}$As barrier is absent. Without the barrier, the $g$-factor values in both QDs become more negative and the step associated with the change in electron localization is significantly less pronounced. This can be understood from the fact that bulk AlAs has a positive electron $g$-factor of $1.52$~\cite{Winkler2003}: contributions from wave-function tails penetrating into the barrier partially compensate the negative $g$-factors of GaAs and InAs. Such an influence of the barrier in a QDM system was reported previously by Doty et al.~\cite{Doty2006}.
We assume that the barrier is positioned slightly closer to the lower QD (by half a lattice constant). Furthermore, the electron wave functions are localized in the upper parts of the respective dots, a consequence of the trumpet-shaped composition profile with enhanced In content near the top. The electron in the lower QD is therefore more sensitive to the presence of the barrier, resulting in a $g$-factor smaller in magnitude than that of the upper QD.
We note, however, that even in the absence of the barrier the two QDs differ in height, which itself introduces a difference between their $g$-factors. In addition, the electric field shifts the carriers within each dot, further modifying the $g$-factor dependence~\cite{Jovanov2011}.

As shown in Fig.~\ref{fig:ge_theory}(a), we also consider intermediate barrier compositions with reduced Al content, namely $x = 0.11$ and $x = 0.22$. The results demonstrate a gradual emergence of the step with increasing Al content. Since at the extreme electric fields the electron states are well localized in each QD, we associate the $g$-factor values at $F = -5$~kV/cm and $F = 25$~kV/cm with the upper and lower QDs, respectively. Their dependence on barrier composition $x$ is shown in Fig.~\ref{fig:ge_theory}(b). Consistent with the previous results, the difference between the $g$-factors grows with increasing Al content, though this effect weakens progressively at larger $x$. This saturation can be attributed to the large band gap of AlAs, which reduces the penetration of the electron wave function into the barrier.

The calculated hole $g$-factor spans the range $0.037$-$0.061$, smaller than the experimental values yet in reasonable agreement with them. The dependence on electric field is shown in Fig.\,\ref{fig:gh_theory} in the appendix. In contrast to the measured $g_\mathrm{h}$, the calculated value increases with increasing electric field (decreasing gate voltage). However, for such small $g$-factor values, uncertainties in the exact QD morphology or in the material parameters may influence the results.

\section{Conclusions}
\label{sec:conclusions}

We have presented a combined experimental and theoretical study of the electron and hole $g$-factors of the negatively charged trion $X^{-}$ in a single InGaAs QDM across the full electric-field tuning range. The electron $g$-factor exhibits a pronounced step-like change at the tunneling resonance, shifting from $g_\mathrm{e} = -0.336\pm 0.008$ to $g_\mathrm{e} = -0.389\pm 0.003$, providing a direct spectroscopic fingerprint of molecular orbital formation, with the difference attributed to the distinct height, composition, strain profile, and AlGaAs barrier proximity of each QD. The hole $g$-factor remains nearly constant at $g_\mathrm{h} \approx 0.094$, exhibiting a weak modulation near the anticrossing voltages due to Coulomb-mediated deformation by the tunneling electron, and neither $g_\mathrm{e}$ nor $g_\mathrm{h}$ shows any measurable magnetic-field dependence. The findings for $g_\mathrm{e}$ are quantitatively reproduced by an eight-band $\mathbf{k}{\cdot}\mathbf{p}$ model combined with a configuration interaction treatment of the trion states, predicting dot $g$-factors of $-0.424$ and $-0.398$ in reasonable agreement with experiment, and further showing that the magnitude of the $g$-factor step can be engineered by tuning the Al content of the tunnel barrier. Our results establish electric-field control of the trion $g$-factors as a practical tool for independently tuning the Zeeman splitting of individual QDs, suppressing pure spin dephasing from $g$-factor mismatches, and opening new pathways towards the deterministic generation of high-fidelity photonic cluster states from coupled QD systems.

\acknowledgments
The authors gratefully acknowledge financial support from the BMBF via the QR.N consortium via Projects No. FKZ 16KIS2197 (J.J.F.), No. 16KIS2200 (A.L.), No. 16KIS2193 (S.R.), and No. 16KIS2206 (D.R.) and Germany's Excellence Strategy (MCQST, EXC-2111, 390814868). M.L., N.A. and J.J.F. acknowledge funding by the Bavarian Hightech Agenda within the Munich Quantum Valley doctoral fellowship program (M.L. and N.A.) and Munich Quantum Valley (J.J.F.). K.G. and J.J.F. acknowledge the DAAD-NAWA for financial support via the center-to-center exchange Grant No. 57754510. The project is cofinanced by the Polish National Agency for Academic Exchange. K. G. acknowledges the financing of the MEEDGARD project funded within the QuantERA II Program that has received funding from the European Union’s Horizon 2020 research and innovation program under Grant Agreement No. 101017733 and National Centre for Research and Development, Poland — project No. QUANTERAII/2/56/MEEDGARD/2024. Created using resources provided by Wroclaw Centre for Networking and Supercomputing (http://wcss.pl.). A.L. acknowledges funding by the QuantERA BMBF EQSOTIC Project No. 16KIS2061 as well as the DFG excellence cluster ML4Q Project No. EXC 2004/1.

\appendix
\section*{Appendix}

\label{app:appendix}

\subsection{Sample structure}
\label{app:sample}

\begin{figure}[ht]
    \centering
    \includegraphics[width=\linewidth]{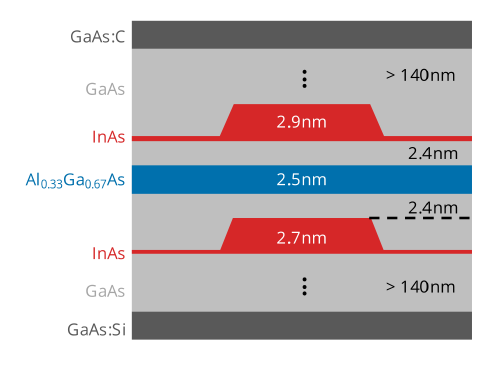}
    \caption{The QDM sample structure. Figure taken from \cite{Lienhart2025}.}
    \label{fig:sample}
\end{figure}

The QDM studied in this work was grown via solid-source molecular beam epitaxy and consists of two vertically aligned indium gallium arsenide (InGaAs) QDs embedded within a gallium arsenide (GaAs) host (see Fig.\,\ref{fig:sample}). The heights of the upper and lower dots were set to 2.9\,nm and 2.7\,nm, respectively, by applying the In-flush procedure during growth \cite{Wasilewski1990}. This particular choice of dot dimensions enables electrical tuning of the conduction-band orbital states (electron states) when the device is driven under reverse bias \cite{Bracker2006}. The upper dot is intentionally made slightly taller to ensure that the hole state remains localized there. The wetting layers of the two dots are separated by 10\,nm. A 2.5\,nm Al$_{0.33}$Ga$_{0.67}$As barrier is inserted between the dots to define the interdot coupling, and an additional 50\,nm Al$_{0.33}$Ga$_{0.67}$As tunneling barrier positioned 5\,nm below the molecule increases the electron escape time.

The QDM is embedded in a p-i-n diode, in which the doped regions provide electrical contacts for gate control. To minimize unintentional carrier injection into the QDM, the diode contacts are placed more than 150\,nm away from the active region. Beneath the diode, an Al(Ga)As/GaAs distributed Bragg reflector is incorporated, and a circular Bragg grating is deterministically aligned with a pre-selected QDM using in-situ electron-beam lithography, enhancing both photon collection and emission \cite{Schall2021}.
Additional details on growth and device fabrication are provided in Refs. \cite{Bopp2023a, BoppMagTun, Lienhart2025, Thalacker2026}.

\subsection{Experimental details}

\subsubsection{Setup}
\label{app:setup}

\begin{figure}[tb]
    \centering
    \includegraphics[width=\linewidth]{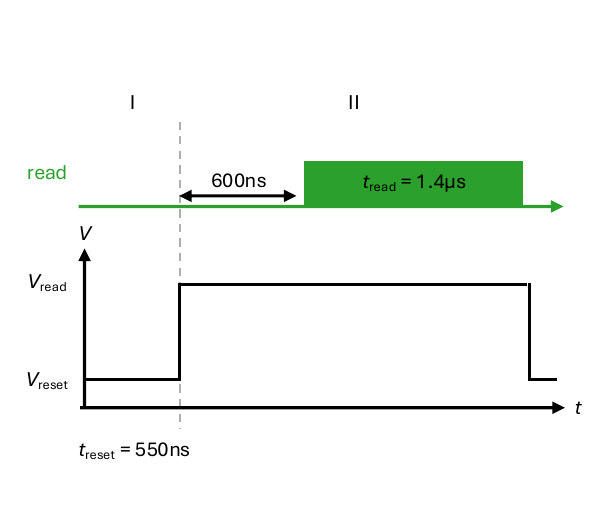}
    \caption{Two phase electrical and optical sequence for determining the electron and hole $g$-factors: in the first phase (reset) a large negative voltage of $V_{\mathrm{reset}}=-10\,$V is applied for $t_{\mathrm{reset}}=550\,$ns. In the second phase (red), a non-resonant laser excites the wetting layer of the QDM for $t_\mathrm{read} = 1.4\,\mu$s at voltage $V_\mathrm{read}$.}
    \label{fig:resetread}
\end{figure}

All measurements discussed in the main text are performed at a temperature of $1.7\,$K using an Attodry2100 dry magnet system with the sample mounted in Voigt geometry.
The $g$-factor measurements are performed at a repetition rate of $390\,$kHz, corresponding to a period of $T = 2.564\,\mu$s.
The measurements make use of a two-phase electrical and optical sequence, shown in Fig.\,\ref{fig:resetread} and described in the following. In the first phase (reset), the QDM is prepared in a zero-charge ground state by applying a large negative voltage of $V_\mathrm{reset} = -10\,$V for $t_\mathrm{reset} = 550\,$ns, efficiently emptying the dot. In the second phase (read), a non-resonant laser excites the wetting layer of the QDM for $t_\mathrm{read} = 1.4\,\mu$s at voltage $V_\mathrm{read}$. To allow the voltage to settle before optical readout, a settling time of approximately $600\,$ns is introduced between the voltage switch and the laser pulse, improving the readout signal quality \cite{boppQuantumDotMolecule2022, Thalacker2026}.

\subsubsection{\textit{g}-factor data analysis}

\begin{figure}[tb]
    \centering
    \includegraphics[width=\linewidth]{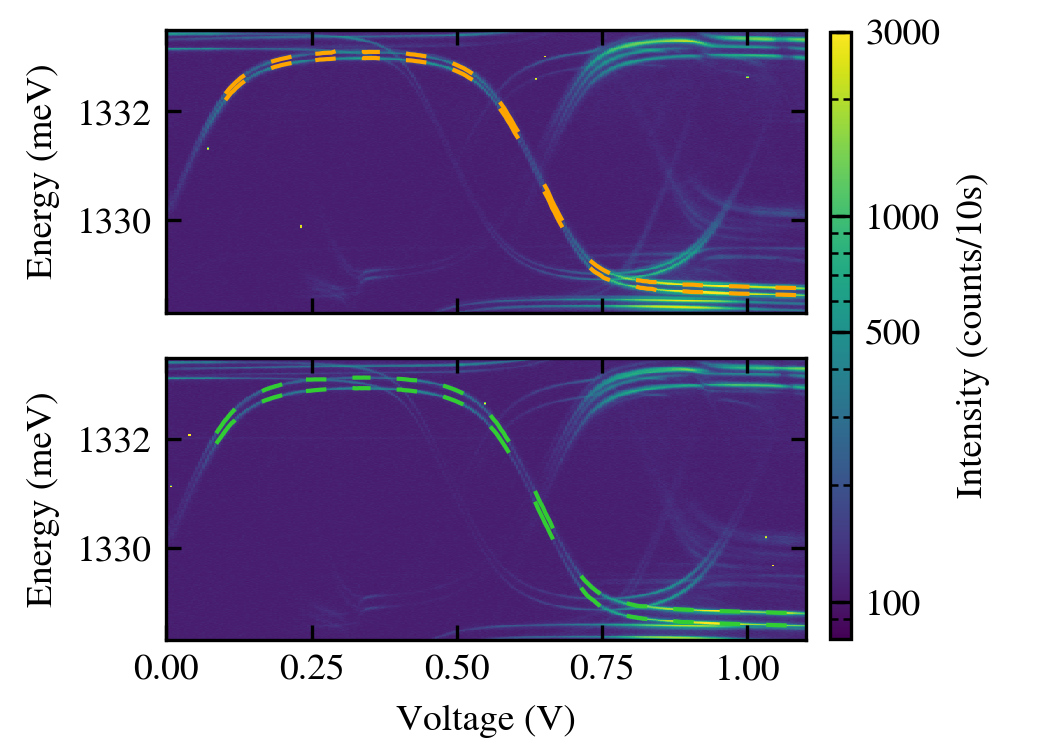}
    \caption{Voltage-dependent photoluminescence measured at $B=8$\,T with the Zeeman-split trion singlet state $X^{-}_{S_0}$ transitions marked with dashed lines. Orange (on top) corresponds to the H-polarized transitions and green (on bottom) corresponds to the V-polarized transitions.}
    \label{fig:PLV_H_V}
\end{figure}

As mentioned in the main text, the electron and hole $g$-factors were obtained via polarization resolved PLVs using wetting layer excitation.
Fig.\,\ref{fig:PLV_H_V} depicts a PLV measurement conducted at $B=8$\,T. The $X^-_{S_0}$ energy lines Zeeman-split by the magnetic field are marked with dashed orange  lines for the H-polarized transitions and dashed green lines for the V-polarized transitions, corresponding to the excerpts displayed in Fig.\,\ref{fig:gfactor_Xm_Voigt}(c). The dashed lines are created by locating the peaks using the Lorentzian fits mentioned in the main text in section \ref{sec:gfactor_experiment}.

\subsubsection{Diamagnetic shift coefficients}

The diamagnetic shift coefficients in the direct configuration (voltages above $V_{e^{-}}$ in Fig.\,\ref{fig:gfactor_Xm_Voigt}(a)) are $(4.2 \pm 0.1)\,\mu\mathrm{eV/T^2}$ and $(2.9 \pm 0.1)\,\mu\mathrm{eV/T^2}$ for the V- and H-polarized branch pairs, respectively. Below $V_{e^{-}}$, the coefficients increase and flatten out near $V_{1,X^{-}}$ around $0.3\,$V, reaching $(6.4 \pm 0.1)\,\mu\mathrm{eV/T^2}$ and $(4.8 \pm 0.1)\,\mu\mathrm{eV/T^2}$, and increase further at even lower voltages.

The increase in the diamagnetic coefficient upon transition from the direct to the indirect exciton configuration reflects the larger electron-hole separation in the indirect state, which corresponds to an effectively larger exciton Bohr radius and therefore a greater sensitivity to the applied magnetic field~\cite{Finley2002}.

\subsection{Theoretical model: calculation details}
\label{app:theory}

\subsubsection{System geometry}

As described in Ref.~\cite{Thalacker2026}, the QDs' geometry is defined via the truncated Gaussians~\cite{Lienhart2025}, where the QD top surfaces are given by
\begin{align*}
    S(x,y) =  \min \qty( w \exp{-\frac{x^2+y^2}{2 d^2}} , h),
\end{align*}
where $h$ is the QD height, $d$ is related to the diameter, and $w$ defines the Gaussian steepness. The values we used for the lower and upper QDs are (the subscript $l$ and $u$, respectively): $h_\mathrm{l} = 3.5 a$, $d_\mathrm{l} = 48 a$, $w_\mathrm{l} = 25a$,  and  $h_\mathrm{u} = 4 a$, $d_\mathrm{u} = 52 a$, $w_\mathrm{u} = 25a$, where $a$ is the lattice constant. Both dots are placed on the $a$ thick wetting layers. The distance between QDs (from the base to the base) is $17.5a$. The Al$_{0.33}$Ga$_{0.67}$As barrier is $4.5a$ thick.

We utilize the trumpet-shape composition profile~\cite{Jovanov2012,Migliorato2002}, where the local indium composition in the dots is given by
\begin{align*}
    C(\bm{r}) = C_\mathrm{b} + \qty(C_\mathrm{t} - C_\mathrm{b}) \exp{ -\frac{\sqrt{x^2+y^2} \exp{-z/z_0}}{r_0}},
\end{align*}
where $C_\mathrm{t}$ and $C_\mathrm{b}$ are the maximum and the minimum of In content in the QD, respectively;  $r_0$, and $z_0$ are parameters defining the profile extension. In both QDs, we took $C_\mathrm{t} = 0.42$ (i.e. In$_{0.42}$Ga$_{0.58}$As), $C_\mathrm{b} = 0.3$, and $r_0 = 20 a$. We took $z_{0\mathrm{l}} = 3.5 a$. and $z_{0\mathrm{u}} = 4 a$, for the lower and the upper QD, respectively. Finally, to account for material intermixing, the resulting composition profile is processed by a Gaussian blur with the standard deviation of $a$. 

\subsubsection{The hole g-factor}

\begin{figure}[b]
    \centering
    \includegraphics[width=\linewidth]{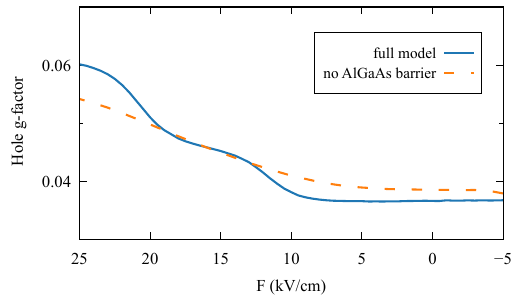}
    \caption{The hole $g$-factor as a function of the electric field.}
    \label{fig:gh_theory}
\end{figure}

The calculated values of the hole g‑factor as a function of the electric field are shown in Fig.~\ref{fig:gh_theory}. Since the hole remains localized in the upper dot for all field values, the influence of the AlGaAs barrier is less pronounced than in the electron case discussed in the main text.

\bibliography{references}

\end{document}